# Two-photon luminescence of single colloidal gold nanorods: revealing the origin of plasmon relaxation in small nanocrystals

Céline Molinaro[1], Yara El Harfouch[1], Etienne Palleau[1], Fabien Eloi[1], Sylvie Marguet[2], Ludovic Douillard[1], Fabrice Charra[1], Céline Fiorini-Debuisschert[1,*]

[1] : SPEC, CEA, CNRS, Université Paris-Saclay, CEA Saclay, 91191 Gif-sur-Yvette cedex, France
[2] : NIMBE, CEA, CNRS, Université Paris-Saclay, CEA Saclay, 91191 Gif-sur-Yvette cedex, France
[*] : corresponding author, e-mail : celine.fiorini@cea.fr

## Abstract

**The two-photon luminescence (TPL) of small 10 nm x 40 nm colloidal gold nanorods (GNR) is investigated at the single object level, combining polarization resolved TPL and simultaneously acquired topography.** A very high dependence of the TPL signal with both the nanorods longitudinal axis and the incident wavelength is observed confirming the plasmonic origin of the signal and pointing the limit of the analogy between GNRs and molecules. The spectral analysis of the TPL evidences two emission bands peaks: in the visible (in direct connection with the gold band structure), and in the infrared. Both bands are observed to vary quadradically with the incident excitation beam but exhibit different polarization properties. The maximum two-photon brightness of a single GNR is measured to be a few millions higher than the two-photon brightness of fluorescein molecules. **We show that the important TPL observed in these small gold nanorods results from resonance effects both at the excitation and emission level : local field enhancement at the longitudinal surface plasmon resonances (LSPR) first results in an increase of the electron-hole generation. Further relaxation of electron-hole pairs then mostly leads to the excitation of the GNR transverse plasmon mode and its subsequent radiative relaxation.**

## Key words (5-6)

Colloidal gold nanorods, field enhancement, electron-hole generation, two-photon luminescence, localized surface plasmons

## Introduction

Since colloidal gold nanoparticles can produce important and highly localized electromagnetic field enhancement effects, they have been the field of intense studies both from a fundamental point of view and also for their possible applications in various fields.[1]

Although metals,[2] and gold[3] more particularly, are known to present very low luminescence quantum yield, a huge luminescence is observed from gold nanostructures excited at their plasmon resonance.[4] For gold nanorods (GNRs) more particularly, quantum yields enhancements over a million[5] have already been demonstrated compared to gold metal. GNRs, now often referred as plasmonic molecules,[6–8] reveals thus as an interesting alternative to molecules for labeling and imaging in biology.[9–12] Their main advantages are that they first present a high photochemical stability and also that their signal does not bleach nor blink.[11]

Due to the rather poor properties of most common biological labels, the case of in-depth imaging and non-linear microscopy is even more demanding.[13] In this respect, together with their quite high 2-photon brightness[14], GNRs also advantageously allow a quite easy tuning of their resonance in the near infrared spectral range. GNRs indeed display two different localized surface plasmon resonances: $\lambda_{LSP}^T$ around 530 nm for the transverse resonance and $\lambda_{LSP}^L$ above 650 nm for the longitudinal resonance, the latter being directly adjustable with the GNR aspect ratio during synthesis.[15] Understanding the origin of TPL in gold nanorods appears thus of fundamental importance towards optimizing their labeling performances.

Although directly correlated to the excitation of longitudinal surface plasmon modes,[16] the detailed origin of the important luminescence in GNRs is still a matter of debates.[17,18] As initially proposed by Sonnichsen et al.,[19] particles plasmons are known to relax through two paths: a direct radiative decay[20] and a non-radiaive decay occurring via the generation of electron-hole pairs, which may also lead to photon emission following radiative recombination. Although the plasmon mode is most of the time shown to shape the one-photon luminescence of GNR,[21–25] the case of two-photon luminescence (TPL) appears more complicated. In some studies, the TPL spectral features reveal the contribution of the band structure of gold, TPL being thus assigned to increased density of states for electron hole recombination near the X and L symmetry points in the gold band structure.[26,27] However, apparently contradictory results were also reported by Knittel et al.[28] considering rods of different crystallinity or shapes: the authors show that the spectral shape of TPL is only determined by the nano-objects geometry and thus their plasmon modes.

The problem appears all the more complex that the luminescence of GNRs seem to depend of a lot of parameters : from the nano-objects crystalline properties to their size or shape[4] and the possible influence of multipolar modes[29] ; with a large number TPL studies devoted to the case of large diameter GNRs.[30,31] More particularly, the case of objects with diameter typically smaller than 10 nm, that exhibit dipolar plasmon properties and for which the absorption cross section $C_{abs}$ is also much higher than the scattering cross section $C_{sca}$, has mainly been restricted to ensemble measurements.[32,33] In solution, however, the GNRs present various and unknown spatial orientations, together with dispersion in size and geometry. In addition, clusters effects might also appear. [34,35]

Excitation conditions seem also to vary quite a lot from one experiment to another, with the use of either continuous-wave or pulsed excitation and the consideration of excitation wavelength close to either the TSPR or LSPR, for which ohmic losses are quite different …

An extensive study is presented in this work to address some of these points. Our aim was to consider in detail the role of both the GNR transverse and longitudinal resonances, one of the other questions being "How far can we go with the analogy between GNRs and plasmonic molecules?".
Considering one photon luminescence, plasmon emission was indeed shown to be comparable to molecular fluorescence, exhibiting an equivalent of the so-called Kasha rule[21]. However, in contrast to molecules whose resonances are governed by localized electronic states called molecular orbitals, the electronic states of metal nanoparticles above 1 nm [36] are collective and delocalized over the whole conducting volume of the particle.

Here, we report the detailed investigation of the two-photon luminescence (TPL) of single 10 nm x40 nm colloidal GNRs. Twenty single nano-objects have been extensively characterized, considering the influence of both the excitation beam polarization and wavelength on their TPL.
We show that unlike the case of molecular transitions from fundamental to first excited state, the GNR transverse resonance (TSPR) cannot be directly excited using the simultaneous absorption of two IR photons. Considering simultaneously acquired topography (using atomic force microscopy – AFM) and polarization resolved optical analyses, we show instead that TPL results from two subsequent steps involving both the longitudinal and the transverse surface plasmon resonances of GNRs: an increase of electron-hole generation first results from local electric field enhancement at the LSPR wavelength ; relaxation of electron-hole pairs through the excitation of the transverse TSPR mode of the rod playing then an important role.

# Results and discussion

40-nm long, 10-nm diameter cylindrical GNRs were synthesized using a now well-established colloidal seed-mediated procedure.[37,38] Transmission electron microscopy (TEM) characterization was performed following GNRs immobilization onto specific substrates: as exemplified in **figure 1**, the perfect, single crystalline nature of the GNRs is clearly evidenced. TEM analysis of numerous isolated GNRs further reveals the very low polydispersity of the GNRs in size and shape, in accordance with the rather small full width at half maximum (FWHM) of the LSPR resonance that can be deduced from the GNR extinction spectrum measured in solution (see figure 4 further below).

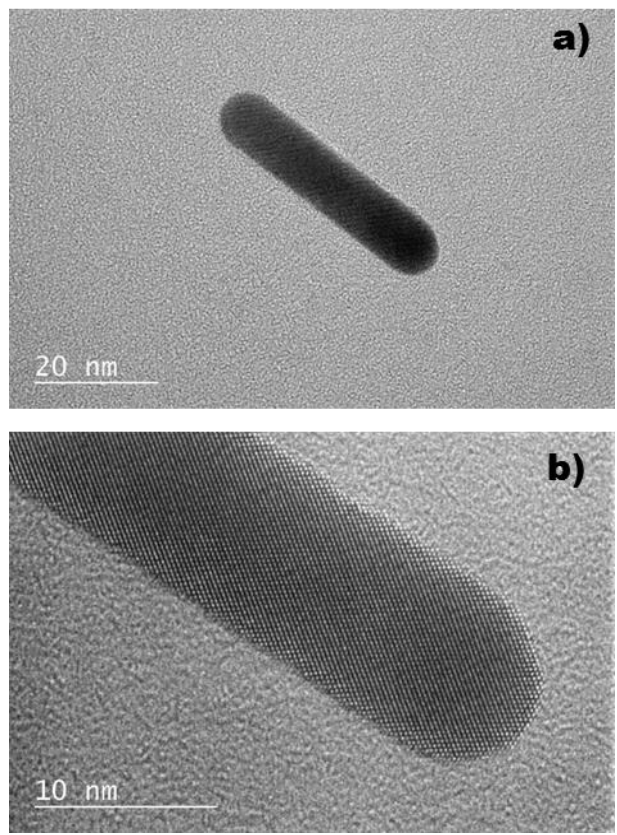

**FIGURE 1 : TEM images of the considered 10 nm x 40 nm gold nanorods (a). High resolution image displaying the crystalline nature of the rod (b)**

Indium Tin Oxide (ITO) coated glass cover slides including e-beam lithography fabricated grids with specific landmarks **(see figure S1 / Supporting info.)** were chosen as substrates in order to enable both AFM and SEM characterization of identical nano-objets **(figure S2 / Supporting info.)**. The slides are treated using UV-ozone in order to optimize the GNRs immobilization though electrostatic coupling between the negatively charged hydroxyl groups that result at the ITO substrates, with the positively charged CTAB (Cetyl trimethylammonium bromide) surfactant naturally present around GNRs following the considered seed-mediated synthesis procedure. The density of immobilized GNRs can quite easily be controlled through the deposition of droplets of solutions with different concentration, nonspecific coupling being eliminated through careful rinsing (ethanol) following drying of the GNRs water solution droplet.

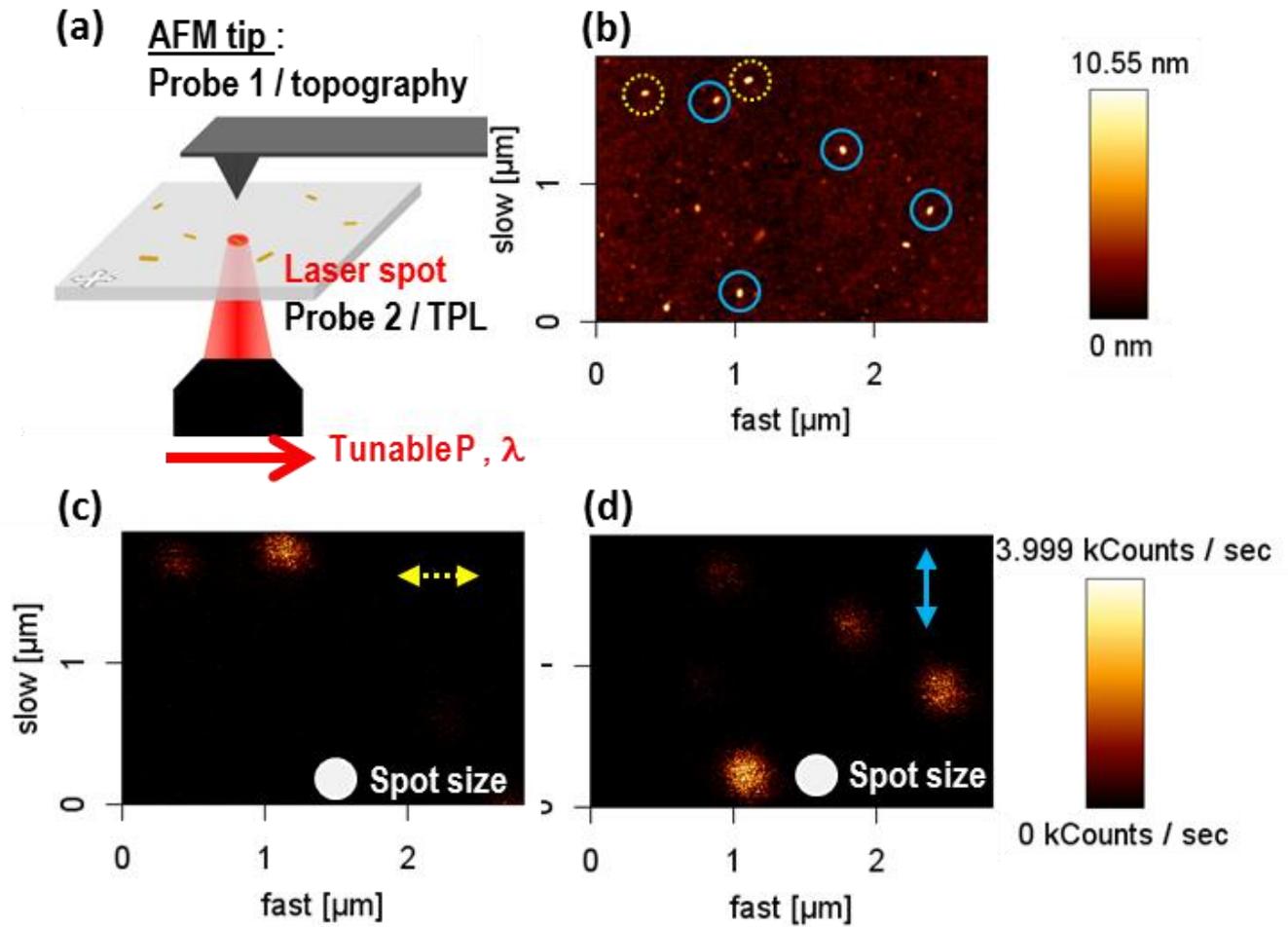

**FIGURE 2 :** (a) Simplified scheme of the experimental set-up. After the preliminary alignment of the AFM tip (topography probe) at the laser spot (TPL probe, beam waist 400 nm, represented by a white disk on the TPL images (c) and (d)), the sample is raster scanned enabling simultaneous topography (tapping mode) and TPL recording (use of an oil immersion microscope objective, NA = 1.3). (b) Tapping mode AFM topography characterization (2 x 3 µm$^2$) of 10 nm x 40 nm gold nanorods immobilized onto ITO glass cover slides (rods which longitudinal axis is mainly oriented along the horizontal direction are spotted by a yellow dotted circle, rods pointing in the perpendicular direction being spotted by a blue circle). (c-d) Simultaneous TPL images (CPM measurements) resulting from the GNRs excitation at $\lambda$ = 800 nm with two perpendicular polarizations : (c) laser polarized along the cantilever (so-called Horizontal polarization – yellow dotted arrow), (d) Vertical polarization (laser polarized perpendicular to the cantilever, continuous blue arrow). The power of the incoming beam at the sample was limited to 20 µW (P=0.5 GW/cm$^2$ at the beam focus) in order to prevent any damage or changes in the GNRs.

The two-photon luminescence (TPL) of single 10 nm x 40 nm colloidal gold nanorods was characterized using a channel plate multiplier (CPM) working in the photon counting mode (Perkin Elmer MP-993-CL) . For this purpose, a specific set-up was developed in order to correlate each GNR

TPL properties with the GNR geometry, orientation and environment. As schematically represented in **figure 2**, each GNR is simultaneously characterized using two probes that are previously aligned (see **figure S3** in the supplementary information for more details): (1) an AFM tip for topography characterization (NanoWizard III, JPK), and (2) a focused laser spot (microscope objective NA = 1.3 leading to a beam waist w = 400 nm) coupled to the sample through an inverted optical configuration, for the two-photon excitation of the nanorods (Tsunami, Spectra Physics – 740 nm < $\lambda$ < 950 nm). The same objective is also used for collection of the luminescence.

The orientation of each GNR can clearly be distinguished even if a slight broadening of the GNRs is observed due to the AFM tip convolution with the rods (figure 2b). Nanorods size and orientation was also confirmed using additional SEM analyses of the same objects (see **figure S3** in the supplementary information). As evidenced in figure 2c and d, a quite important signal is obtained that appears to be highly dependent on the GNRs orientation for a given incoming laser beam polarization. Further angular dependence studies have been realized on different single GNRs with different spatial orientations. As represented in figure 3, fitting of the experimental results leads to a $\cos^4(\theta)$ dependence, with $\theta$ the angle between the laser polarization and the rod longitudinal axis, which indicates not only the dipolar origin of the signal but evidences also the quadratic dependence of the signal with the excitation beam. This experiment was repeated and reproduced on more than twenty GNRs. The correlation between topography measurements and TPL intensity points out that the angle of maximum luminescence of a GNR is observed each time in the direction of its longitudinal axis. The average deviation between the angle of maximum TPL and the spatial orientation of GNRs is less than 3°. So, unlike the case of molecular fluorescence, where the same transition can be addressed either using the absorption of one visible photon or two IR photons, there can be no TPL resulting from a two-photon excitation process along the transverse axis of the nanorods.

Figure 3b evidences the huge ratio between maximum and minimum TPL (a few hundreds). Although quite small, the minimal signal measured for an excitation polarization along the GNR transverse axis is clearly distinguished from noise, varying moreover quadratically with the excitation beam intensity as expected from the two-photon nature of the luminescence. Interestingly, as detailed in the supplementary information, the maximum two-photon brightness recorded for one single GNR was measured to be quite large, corresponding to the two-photon brightness of a few millions fluorescein molecules excited in the same conditions.

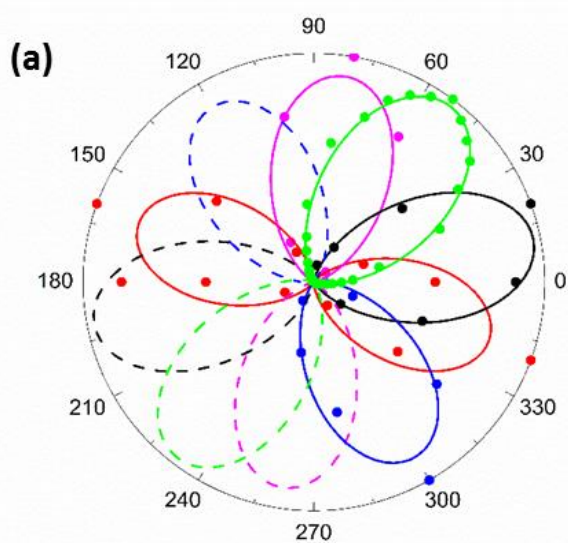
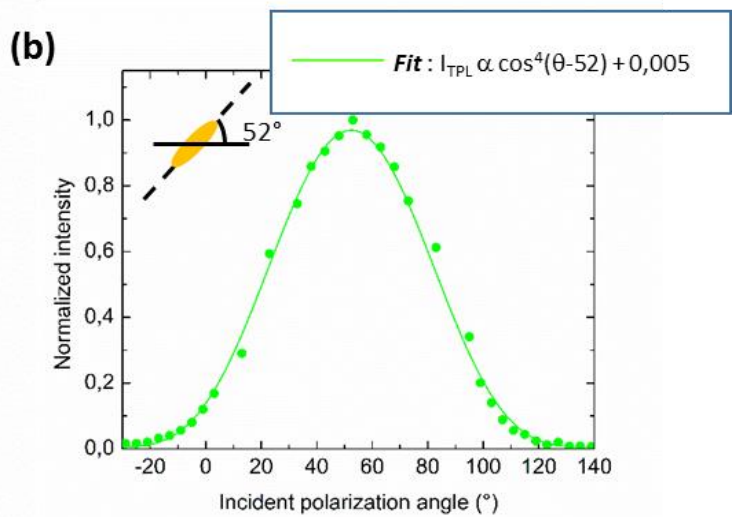

**FIGURE 3 : (a) :** Polar representation of the dependence of the TPL intensity of different single GNRs with the incident polarization angle θ (excitation wavelength : λ = 800 nm). Dots correspond to the experimental data and lines to a $\cos^4(\theta)$ dependence. The case of 5 different GNRs is here represented (even if more than 20 nano-objects were considered showing the same dependence). As represented in (b), an increased number of measurements was performed in the case of one object showing the very good agreement between the experimental data and fit, the only adjustable parameter being the constant a, corresponding to the signal at minimum (out of resonance signal).

The influence of the exciting beam wavelength $\lambda_{exc}$ was further analyzed considering a polarization along the longitudinal axis of the GNR: as represented in figure 4, the maximum TPL is obtained at the GNR longitudinal surface plasmon resonance ($\lambda_{exc} = \lambda_{LSP}^L$), in agreement with previously reported studies[12,33]. The TPL data measured for a single GNR appear moreover to be very well described (no external adjustable parameters) by the theoretical dependence that can be derived from the Gans-Mie model, the most famous quasistatic theory for calculating the light scattering and absorption of small nanocrystals presenting an ellipsoidal shape.[37] The absorption ($C_{abs}$) and scattering cross sections ($C_{sca}$) can more particularly be written :

$$C_{abs} = \frac{2\pi}{3\lambda}\varepsilon_m^{3/2}V \sum_i \frac{\varepsilon_2/(n^{(i)})^2}{\left(\varepsilon_1+[(1-n^{(i)})/n^{(i)}]\varepsilon_m\right)^2+\varepsilon_2^2} \quad (1)$$

$$C_{sca} = \frac{8\pi^3}{3\lambda^4}\varepsilon_m^2 V^2 \sum_i \frac{(\varepsilon_1-\varepsilon_m)^2+\varepsilon_2^2/(n^{(i)})^2}{\left(\varepsilon_1+[(1-n^{(i)})/n^{(i)}]\varepsilon_m\right)^2+\varepsilon_2^2} \quad (2)$$

$$C_{ext} = C_{abs} + C_{sca} \quad (3)$$

where $\varepsilon_m$ is the dielectric constant of the surrounding medium, ε the dielectric constant of the metal with $\varepsilon = \varepsilon_1 + i\varepsilon_2$. λ is the wavelength of the incident light and $n^{(i)}$ the depolarization factor described by:

$$n^{(a)} = \frac{2}{R^2-1}\left(\frac{R}{2\sqrt{R^2-1}}\ln\left(\frac{R+\sqrt{R^2-1}}{R-\sqrt{R^2-1}}\right) - 1\right) \quad (4)$$

$$n^{(b)} = n^{(c)} = (1-n^{(a)})/2 \quad (5)$$

Interestingly, although the width of the LSP resonance in water is observed to be larger due to polydispersity, the maximum TPL of most single GNRs is found to be at the same wavelength as the longitudinal LSPR determined from the extinction spectrum of a GNRs solution in water. As confirmed from the extinction spectrum measured for an ensemble of particles lying onto ITO (see **figure S4 / Supporting info.**), this is due to the fact that the refractive index of the surrounding medium for a GNR in water ($n_m^{water}$ = 1.33) happens to be quite close to the refractive index that can be considered for a GNR immobilized at an ITO-air interface ($n_m^{ITO-air}$ ≈ (1+1.6)/2=1.3, with 1 and 1,6 being respectively the indices of air and ITO at the considered wavelengths).

Figure 4 shows more particularly that the TPL closely follows the dependence of the NPs absorption determined from the Gans theory: $I_{TPL}$ is directly proportional to $(C_{abs})^2$, with $C_{abs}$ the GNR absorption cross section (please note that for such small GNR, $C_{scatt}$ ≈ 3 nm² appear to be more than two order of magnitude less that $C_{abs}$ ≈ 1800 nm² (see **figure S5 / Supporting info.**). We conclude that increased

TPL in single GNRs directly originates from increased nonlinear absorption at the longitudinal plasmon resonance.

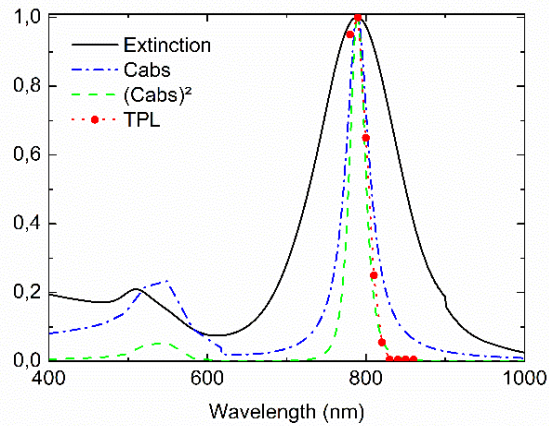

**FIGURE 4: TPL excitation spectrum recorded for a single 10 nm x 40 nm GNR immobilized onto ITO and excited along its longitudinal axis (red dots). The extinction spectrum of a water solution (black line – *ensemble measurements*) of identical GNRs is shown for comparison. The dotted dashed blue and dashed green lines are the dependence of the GNR absorption cross section $C_{abs}$ and $C_{abs}^2$, respectively, as deduced from the Gans theory (normalized data - see figure S5 for quantitative data).**

Quite interesting, although most studied GNRs were found to present a TPL maximum at the resonance value $\lambda_{LSP}^L$ deduced from the theoretical Gans model, few GNR (around 5 to 10%) were also found to present a maximum TPL slightly shifted from $\lambda_{LSP}^L$ : this reveals the slight GNRs polydispersity and is in good agreement with the larger resonance observed in solution.

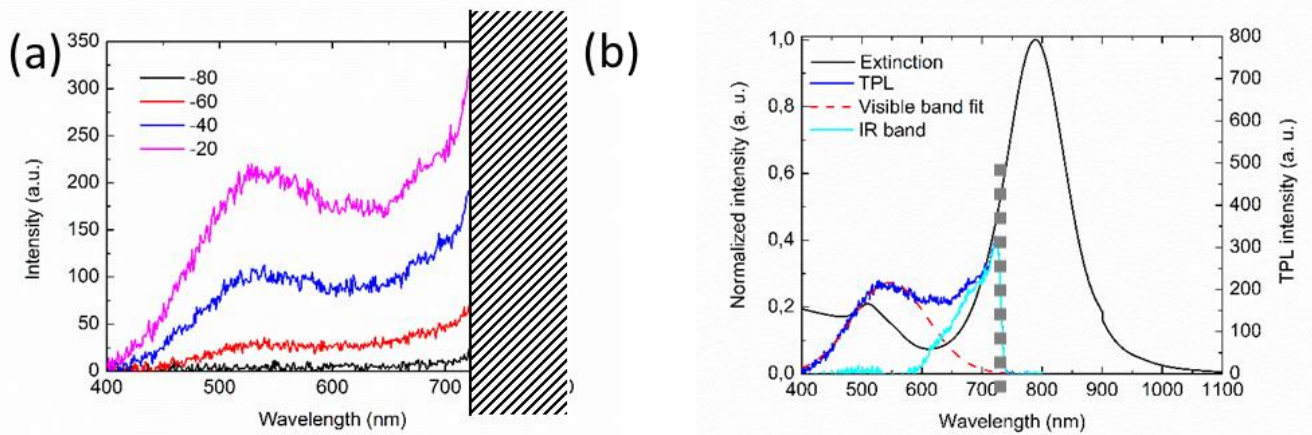

**FIGURE 5: (a) TPL emission spectra of a single GNR excited at $\lambda$ = 800 nm, with its long axis oriented at -20° for different incident exciting beam polarizations. GNRs were studied at their resonance wavelength ($\lambda$ = $\lambda_{LSP}^L$ ) with a constant power of 20 µW. The spectra are corrected from the wavelength sensitivity of our CCD camera (Andor DU401-BR-DD). (b) Superposition of a single GNR emission spectrum (blue line) with the normalized extinction spectrum of GNRs in solution (black line). The dotted grey line represent the wavelength above which filters are used to cut the exciting laser beam (hatched area in (a)). The visible part of the TPL is fitted using a Gaussian dependence (red dashed line), the IR part of the TPL being deduced after subtracting this fitted visible band to the full experimental emission spectrum (cyan line). Applying this**

**treatment to all the curves presented in (a) shows that the relative amplitudes of the visible and IR contribution appears to be independent of the exciting beam polarization.**

As a complementary characterization, the spectral content of the TPL signal was also analyzed through a spectrometer coupled to a CCD camera (Andor DU401-BR-DD – see supplementary information). As shown in **figure 5**, two emission bands are evidenced, varying both quadratically with the laser excitation power and peaking, one around $\lambda_{TPL}^{max, VIS}$ = 547 ± 4 nm and the second one in the IR (see figure 5) as already reported by other authors at the level of ensemble measurement in solution.[40] The maximum of this second peak is however impossible to determine directly from the experiment due to the use of short pass filters to block the excitation laser (average OD >> 10 for wavelengths above 750 nm, see details in **S3**). The two bands contributions can be separated through fitting of the visible emission band and further determination of the remaining contribution centered in the IR (possibly around $\lambda$=800 nm, the limited data in this spectral range prevents any accurate fit). As shown in figure 5b, the visible contribution of the TPL spectrum appears to be centered very close to the GNRs transverse surface plasmon band. Considering that the maximum TPL is obtained at $\lambda = \lambda_{LSP}^L$ and for an exciting beam polarization along the longitudinal axis of the GNRs, there can be no direct excitation of the perpendicular GNR transverse plasmon mode. According to this, TPL can only result from the excitation of electron-hole pairs, which relaxation can then either excite other GNRs plasmon modes or directly recombine radiatively. Indeed, the maximum emission wavelength of the visible TPL contribution $I_{TPL}^{max, VIS}$ happens also to be very close to interband transition at the L-symmetry point of the first Brillouin zone of gold, in good agreement with the crystalline structure of the nanorods.[41,42] Finally, it is important to note that the spectral position of these two bands together with their relative intensities were observed to be independent on the excitation wavelength, ruling out any Raman origin of the signal.[43,44]

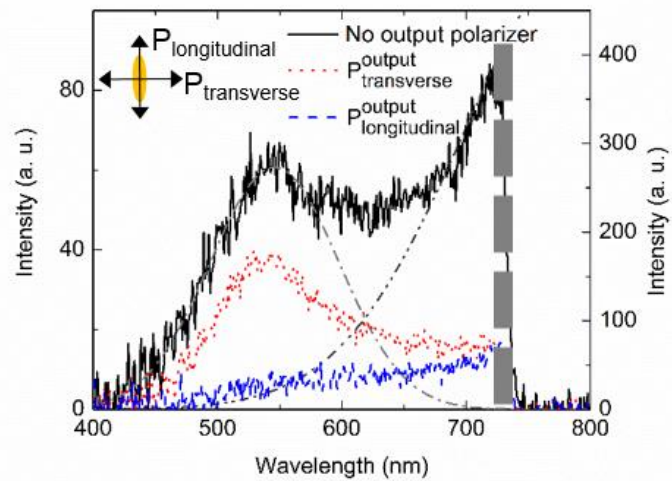

**FIGURE 6: Polarization resolved analysis of the spectrum of a single GNR TPL emission. The GNR was excited at its resonance wavelength with an incident power of 20 µW, and an exciting beam polarization along its longitudinal axis (for maximum TPL). The black line corresponds to the full emission spectrum as measured in the absence of any polarizer at the signal output (left scale). The dashed gray lines correspond to both the so-called visible or IR contribution of the TPL (see discussion above in the text). The bold dotted grey line represent the wavelength above which filters are used to cut the exciting laser beam. The blue and red lines correspond to the TPL spectrum measured with an output polarizer along the GNR longitudinal or transverse axis, respectively (right scale).**

The polarization properties of the TPL of different single GNR were further investigated in detail in order to distinguish between either a direct radiative recombination of electrons and holes or an emission occurring via the GNR plasmon modes. For this purpose, a polarizer was added at the microscope output. Single GNR are excited with an incident beam at the resonance wavelength and with an incident polarization along their longitudinal axis. Variation of the output polarizer angle directly enables the measurement of the TPL spectrum emitted either along the GNR transverse or longitudinal axes. Spectra are shown in **figure 6**.

As evidenced in figure 6, the two so-called visible and IR bands that contribute to the TPL emission, appear to be quite differently polarized: the band peaking around $\lambda$ = 550 nm is 80% polarized along the GNR transverse axis, i.e. a direction perpendicular to the exciting beam polarization thus confirming the intermediate production of electron and holes. In contrast, the IR band appears to be mainly unpolarized with nearly equal signals in both polarization directions (polarization degree less than 10% along the GNR longitudinal axis), as already observed in the case larger gold nanoantenna considering however the whole TPL emission spectrum[45].

The visible TPL emission band appears thus quite strongly connected to the GNR transverse plasmon resonance, thus leading to the mechanism illustrated in figure 7 where TPL is described by a 4-step mechanism[46]:
**(1)** First two photons excite an electron from the d-band to the unoccupied states in the sp-band, this 1$^{st}$ step benefiting from field enhancement at the GNR LSPR[31,47,48]. **(2)** As a second step, hot electrons in the sp-band move close to the Fermi energy level by interband scattering processes. Then **(3)** electron and hole can recombine radiatively by interband relaxation near X and L symmetry points of the first Brillouin zone. In this case, emission is only shaped by the particles crystallinity and no specific polarization properties are expected. Alternatively **(4)**, a non-radiative decay of electrons in sp-band can induce an excitation of plasmons, which finally decay radiatively giving photons with specific polarization properties and energies close to their plasmonic modes.

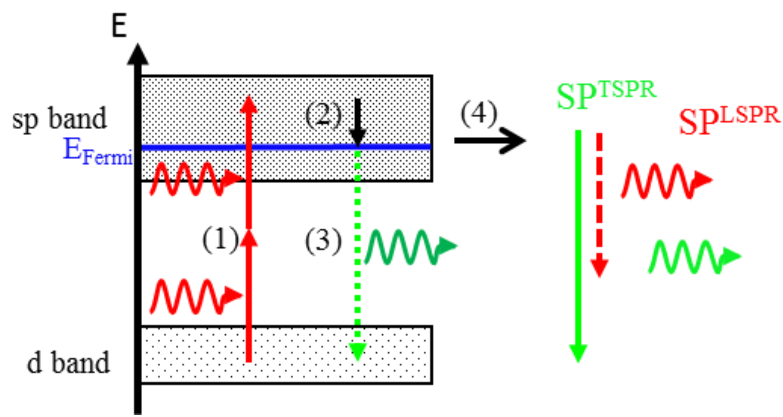

**FIGURE 7: Schematic diagram of the proposed TPL mechanism : (1) two-photon interband absorption (d-band to sp-band) creates hot electron-hole pairs (plasmon enhanced absorption at the longitudinal plasmon resonance) - (2) Thermalization: Electrons in the sp-band move closed to the Fermi energy level - (3) Radiative recombination of electron/hole pairs -(4) Decay of electron hole pairs into the excitation of either longitudinal (LSPR) or transverse (TSPR) plasmon modes that subsequently radiate.**

As evidenced in figure 6, this latter plasmon-mediated process appears to be the major process at the origin of the visible TPL emission of the GNRs considered in our studies. The case of the IR emission band is less clear, one of the problem being that the collection of this band is only limited to a very small spectral area due to the filters used to cut the exciting laser beams … Further studies would be necessary to clarify this point.

# Conclusion

Following polarization resolved single GNR TPL studies, we have evidenced here that the two mechanisms regularly opposed for the luminescence of plasmon particles can clearly be discriminated. In other words, both radiative and non-radiative decay processes coexist for the considered small gold nanorods (10 nm x40 nm); the case of longer nanowires being apparently different.[49]

Considering the influence of both the polarization and wavelength of the exciting laser beam shows more particularly that the TPL of small nanorods results from increased absorption processes at their longitudinal plasmon resonance, obviously leading to the creation of electron-holes pairs. Interestingly, unlike in molecular two-photon fluorescence processes, there can be no direct TPL excitation of a GNR through excitation along its transverse axis. At the same time, careful analysis of the polarization properties of single GNR visible TPL, leads to the conclusion that electron-hole pairs mainly relax through the excitation of the GNR transverse plasmon mode. Direct radiative relaxation of electron and holes is also observed, but in a much less extent, which explains the fact that visible TPL is not fully polarized (80% degree of polarization).

The large two-photon luminescence in gold nanorods results thus from the interplay of their two different plasmon resonances (transverse, $\lambda_{LSP}^T$, and longitudinal, $\lambda_{LSP}^L$, plasmon modes) leading to both plasmon enhanced absorption (for an exciting wavelength $\lambda_{exc} \approx \lambda_{LSP}^L$) and plasmon enhanced emission processes ($\lambda_{em} \approx \lambda_{LSP}^T$). Next step will now consist in better understanding the parameters controlling the radiative relaxation processes of electron-hole pairs : above thermalization processes which were recently considered in detail[17], a better understanding is needed regarding the way to control either step 3 and 4 as described in figure 7. For this purpose, further experiments are currently under way, considering single gold nanorods presenting similar plasmon resonances but different volumes.

# Acknowledgements
The authors would like to thank the C'nano 2010 program "INPACT", ANR Blanc 2012 project "HAPPLE" and CEA "Programme transverse NanoScience" (YEH) for financial support of this work, Jeremie BEAL (UTT) for the realization of landmarks using e-beam lithography and the IS2M laboratoty (*Dr Wajdi Heni, Dr Olivier Soppera*) for the high resolution TEM images of the rod.

# Table of contents

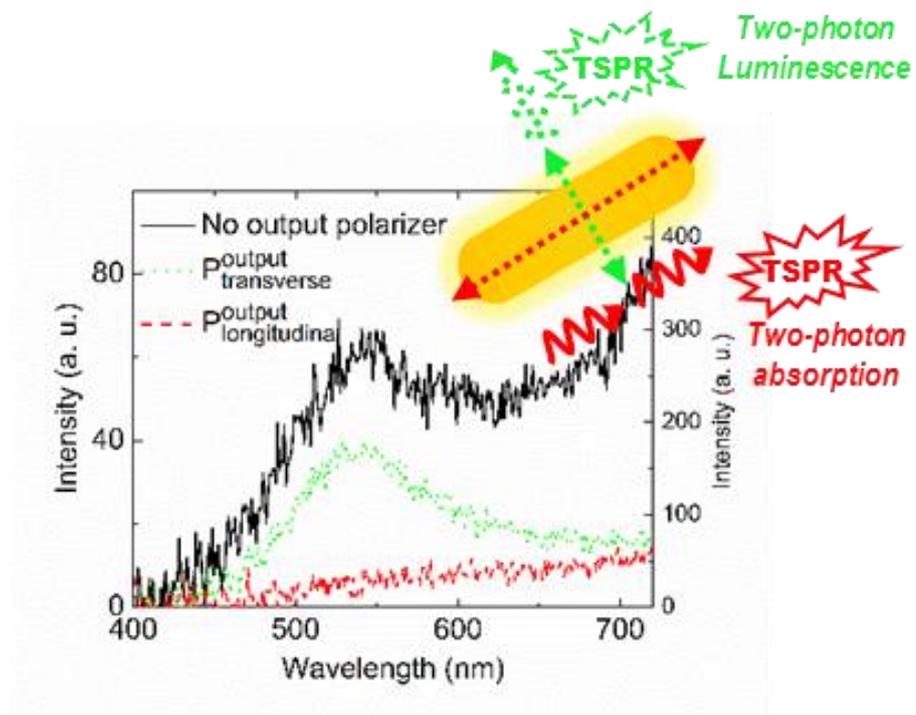